\documentclass{article}
\usepackage{spconf,amsmath,graphicx}
\usepackage{multirow}
\usepackage{multicol}
\usepackage{algpseudocode}
\usepackage{url}
\usepackage{algorithm}
\usepackage{subcaption}
\usepackage{pgfplots}
\pgfplotsset{compat=1.17}

\title{DDSP-SFX: Acoustically-guided sound effects generation with Differentiable Digital Signal Processing}
\name{Yunyi Liu$^{1,2*}$, Craig Jin$^{1}$, David Gunawan$^{2}$\thanks{$^{*}$This research was done while interning at Dolby Laboratory.}}
\address{$^{1}$University of Sydney, $^{2}$Dolby Laboratory}

\begin{document}
%
\maketitle
\begin{abstract}

Controlling the variations of sound effects using neural audio synthesis models has been a difficult task. Differentiable digital signal processing (DDSP) provides a lightweight solution that achieves high-quality sound synthesis while enabling deterministic acoustic attribute control by incorporating pre-processed audio features and digital synthesizers. In this research, we introduce DDSP-SFX, a model based on the DDSP architecture capable of synthesizing high-quality sound effects while enabling users to control the timbre variations easily. We propose a transient modelling technique with higher objective evaluation scores and subjective ratings over impulsive signals (footsteps, gunshots). We propose a simple method that achieves timbre variation control while also allowing deterministic attribute control. We further qualitatively show the timbre transfer performance using voice as the guiding sound. 
\end{abstract}
\section{Introduction}
\label{sec:intro}

Sound effects refer to natural or synthetic sounds different from speech or music. They play an important part in digital media such as film and games, but typically require intensive labour in recording. In recent years, deep generative models have demonstrated their synthesis capability in the audio domain, especially for modelling speech~\cite{samplernn} and music~\cite{nsynth}. However, neural audio synthesis (NAS) methods typically model the audio waveforms or time-frequency representations directly and require a large amount of data and computation power. To model sound effects with varying acoustic characteristics a large, even vast, set of descriptive labels~\cite{Neural-footstep, liu_audioldm_2023}, may be required which is often impractical. 

Sound effects are diverse and usually difficult to describe using words. To this end, many synthesis algorithms focus on using vocalizations to guide sound generation because voice can be an intuitive control interface for end-users. The supervised phoneme-based control-synthesis approach which converts human-produced phonemes directly to sound effects has been widely explored~\cite{kwon_voice-driven_2012, okamoto_onoma--wave_2022, suzuki_speak_2022, takizawa_synthesis_2023}. However, such methods typically require ground-truth labelling between the phonemes and the corresponding sound effects. The unsupervised approach, which entails extracting acoustic features from the guiding sounds, and applying such features to the generated sound effects has also proven to be very effective in musical sound modelling. For example, differentiable digital signal processing~\cite{DDSP} is a popular NAS architecture introduced to take advantage of pre-built digital synthesizers for waveform generation. Neural networks are then used to estimate the parameters for the corresponding synthesizers conditioned on pre-processed acoustic features. Owing to the modular approach and conditioned features, DDSP is lightweight to train and use, while offering controllable audio synthesis over pre-defined audio features such as pitch and loudness. Although Lundberg~\cite{lundberg_data-driven_nodate} has applied DDSP to model motor engine sounds, there have been few attempts at modelling other sound effects especially impulsive sounds such as footsteps or gunshots, two commonly used sound effects in games. Additionally, it remains a critical issue as to how to effectively control the subtle timbre variations of a particular sound effect driven by vocalizations given the differing acoustic characteristics between voice and various desired sound effects. 

In this research, we explore sound effects modelling driven by acoustic features of target sounds utilizing the DDSP architecture. We are particularly interested in using voice to control the subtle timbre characteristics of the target sounds, such as varying the type of engines for a motor sound with fixed pitch and loudness. Novel contributions in this work are that we incorporate a transient modelling method to improve the synthesis of impulsive sound effects and introduce a new approach for time-varying timbre control given a limited dataset. 

\section{Proposed method}

\begin{figure*}[h]
  \centering
  \centerline{\includegraphics[width=\textwidth]{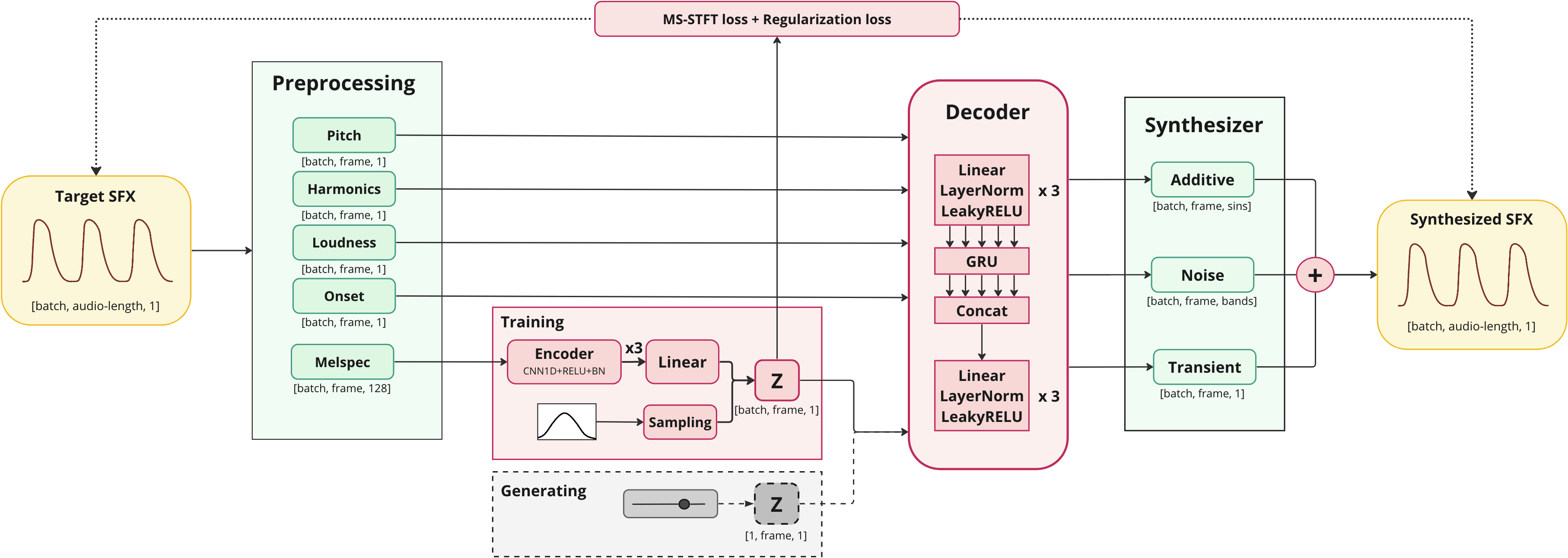}}
  \caption{Proposed model architecture}
  \label{fig: model}
\end{figure*}

We base our model on the Pytorch DDSP implementation by Acids\footnote{\url{https://github.com/acids-ircam/ddsp_pytorch}} and denote our method as DDSP-SFX. Our complete model architecture is depicted in Figure~\ref{fig: model}.

\subsection{Transient modelling}
\label{sec:Transient}

A transient signal (e.g., gunshots, footsteps) has a sharp attack and short sustain, which can be difficult for sinusoidal modelling or subtractive noise modelling~\cite{serra_spectral_1990}. To synthesize transient signals, we employ a similar modelling approach as~\cite{lundberg_data-driven_nodate}, i.e., synthesizing sinusoids in the discrete cosine domain and converting to the time domain using an inverse Discrete Cosine Transform (IDCT). Due to the nature of time-frequency transforms, this results in impulses in the time domain~\cite{verma_analysissynthesis_1998}. For each of the 400 time frames of the 4 s long signal, a decoder network will learn and output the parameters of the sinusoids used for transient modelling: i.e., the frequency $F_n$ and amplitude $A_n$. If no transient is present in a given time frame, then $A_n$ should be $0$. This provides a convenient means for controlling the timbre of the transient signals. The transient modelling equation is defined as:
\begin{equation}
    x[n] = \mathrm{IDCT}(A_{n}*sin(2\pi F_{n}k))
\end{equation}
where $x[n]$ is a transient signal with a window length of 160 samples. We synthesize transient signals for each of the N=400 time frames.  Thus, one of the inputs to the decoder network is an onset amplitude vector of length 400 which indicates the presence or absence of a transient and its amplitude for each of the 400 time frames of the signal. The onset amplitude vector is constructed from the signal spectrogram using a margin-based Harmonic Percussive Source Separation (HPSS) method~\cite{Driedger2014ExtendingHS} with a conservative and large margin parameter value of $8$.  Local peak estimation is performed to obtain the onset amplitude vector where the percussive events take place.\\

\subsection{Harmonic Indicator}
\label{sec:Harmonic}
As the vanilla DDSP synthesizer assumes modelled sounds are harmonic~\cite{noisebandnet}, it will synthesize sinusoids even when the modelled sounds are inharmonic and noisy, thus resulting in artifacts. Therefore, we employ a harmonic indicator vector $H$ to determine the degree of harmonic components present in the sound and thus attenuate the harmonic synthesizer during percussive sounds. To achieve this, during pre-processing we extract a confidence score $C$ (0-100\%) for the pitch estimation of the modelled sound using CREPE~\cite{CREPE} and then apply a sigmoid function to calculate a harmonic indicator value, $H$, as shown below. 
\begin{equation}
    H = 1 / (1 + e^{(-10 (C - 0.7))}) \, \text{.}
\end{equation}
As shown, we found a confidence score above 0.7 is able to achieve a reasonable harmonic indicator value.

\subsection{Timbre encoding}
\label{sec:Timbre}

In addition to controlling the deterministic audio features such as pitch and loudness, we wish to control the timbre variations of the generated sound expressively. To this end, we employ a similar encoder structure introduced by Devis et al.~\cite{devis_continuous_2023} and train DDSP as a Variational Autoencoder (VAE)~\cite{VAE}. We first compute the mel-spectrograms (128 mel-frequency bands, 400 time frames) of our input sounds in the pre-processing stage. They become the input to our encoder, comprising three stacks of convolutional 1-D layers, RELU activation, and batch normalization layers followed by a linear layer. The mean and log-variance output from the encoder are then reparameterized to produce the sampled latent vector $z$: 
\begin{equation}
    z = \mu + \epsilon \cdot {\sigma}^2 \, \text{,}
\end{equation}
where $\epsilon$ is a random value sampled from a unit-Gaussian distribution, $\mu$ is the mean output from the encoder, $\sigma = e^{0.5*\text{logvar}}$, and $\text{logvar}$ is the logarithm of the variance. In this way, we obtain a continuous one-dimensional latent vector along the time axis. The distribution of $z$ is regularized to be close to a unit-Gaussian $\mathcal{N}(0,1)$. After the model has been trained, we can deliberately modify the value of $z$ within $\pm3$ as 99.7\% of the values are within three standard deviations. The most "typical" timbres appearing in the data set are encoded with $z$ close to 0, whereas "rare" timbres are encoded with $z$~values farther away from 0. In the synthesis application stage, we create a control variable within the same range $\pm3$ to replace $z$ for the decoder to synthesize different sound timbres. A global slider is used to adjust this pseudo latent variable and thus vary the timber of the output sound.

\subsection{Loss Function}
Our loss function contains a regularization loss component and a reconstruction loss component. We use the same multi-scale STFT loss as used in DDSP for our reconstruction loss (FFT sizes: 2048, 1024, 512, 256, 128, 64). For the regularization loss, we apply a scaling variable $\beta$ on the regularization loss term to prevent the reconstruction loss from overtaking the total loss~\cite{beta-vae}. The total loss function in our model becomes:

\begin{equation}
\label{eq: loss}
    \mathcal{L} = \mathcal{L}_{rec} + \beta*\mathcal{L}_{reg}
\end{equation}
where the $\mathcal{L}$ indicates the total loss, $\mathcal{L}_{rec}$ is the reconstruction loss, $\beta$ is a scaling variable and $\mathcal{L}_{reg}$ is the regularization loss.

\section{Experiments}

\subsection{Dataset}
We use a publicly available sound effects dataset~\cite{Choi_arXiv2023_01}. It includes 7 categories of sound effects, all of which are sampled at 22.5~kHz and have a length of around 4~seconds. The data used for all of the experiments described in this paper consists only of footsteps, gunshots and motor sounds. We refer to these three sound types as our three data sets. All sounds are trimmed to 4~seconds duration and sampled at 16~kHz. We split the data 90\% for training and 10\% for testing. 

\subsection{Training}
We train our model and the vanilla DDSP model on the same three data sets separately, resulting in three trained models for both the vanilla DDSP and our proposed method. Each model is trained with a batch size of 16 on an RTX 6000 GPU for exactly 100,000 training steps. We chose a size of 1024 for the recursive hidden units of DDSP, 100 sinusoids for the harmonic synthesizer, 100 bands for the noise synthesizer, and a frame size of 160 samples. We use an ADAM~\cite{ADAM} optimizer with a starting learning rate of $1e^{-4}$ which gradually decayed to $1e^{-5}$ after 80\% of the steps. To maintain a stable regularization loss and balance it with the reconstruction loss, we initialize $\beta$ with 0, and activate it only after 10\% of the training steps with $\beta=1$, and then scale it linearly to $1e^3$ until reaching 80\% of the training steps.



\subsection{Evaluation}
To show that our latent space is capable of encoding timbre information of different sound effects, we performed a small-scale listening test on 26 participants. We recorded three human vocalizations emulating the sound effect for each category of our data set as our out-of-domain guiding sounds. We then use the extracted acoustic features from these guiding sounds to perform timbre transfer. For our latent vector, we manually set $z$ as 0, 0.1, 0.5, 1, 2, and 3 for each sound clip and then generate the sound effects correspondingly. We use $z=0$ as our reference track and ask the participants to do a forced comparison test for whether the individual tracks with $z$ set as different values sound identical to or different from the reference. The result is shown in Section~\ref{sec: Timbre encoding}.

We evaluate our model in terms of synthesis quality. The timbre encoding performance is demonstrated qualitatively. For each model, we pair a reference sound with the generated sound. Each model is tasked to synthesize similar waveforms to the reference by taking in the extracted acoustic features from the reference.  We compare the synthesis performance through a series of objective metrics and a subjective listening test. We conduct a second subjective listening test to understand the effectiveness of our timbre encoding. %

\subsubsection{Synthesis performance}

\textbf{Statistical similarity}. Frechet Audio Distance (FAD)~\cite{FAD} is an audio quality evaluation method that compares the feature representations through an embedding layer of a pre-trained audio classification model between the generated sounds and the reference sounds. We compute the FAD score with the VGG model~\footnote{\url{https://github.com/gudgud96/frechet-audio-distance}} to understand the statistical similarity of our synthesized sounds compared with the reference sounds.\\
\textbf{Spectral similarity}. We report the log-spectral distance using a multi-scale STFT to measure the spectral similarity between the synthesized sound and the reference sounds. We use the same window length, hop length and FFT size as we did in the preprocessing to compute the spectrograms. \\
\textbf{Audio quality.} We conduct a listening test to evaluate the generated sound quality. We randomly selected ten examples per category. For each question, we asked the participants to give an absolute category rating for each soundtrack from 1(bad) to 5(excellent). We aggregate our results per category of sounds, meaning that we average the 10-question results and obtain the variance for the aggregated data.

\section{Results}
\label{sec:majhead}

\subsection{Audio similarity}

\begin{table}[htb]
\centering
\begin{tabular}{c|c|c|c}

\hline
\multirow{4}{*Categories}& Footstep & Gunshot & Motor \\
\hline
\hline
Metric & \multicolumn{3}{c}{FAD $\downarrow$} \\
\hline
DDSP & 5.356 & 5.213 & \textbf{6.652}\\ 
DDSP-SFX & \textbf{1.529} & \textbf{2.004} & 7.150\\ 
\hline
\hline
Metric & \multicolumn{3}{c}{Log Spectral Distance $\downarrow$} \\
\hline
DDSP & 0.114 & 0.585 & \textbf{0.177}\\ 
DDSP-SFX & \textbf{0.103} & \textbf{0.446} & 0.182\\ 
\hline
\hline
Metric & \multicolumn{3}{c}{Multi-scale STFT $\downarrow$} \\
\hline
DDSP & 1.63 $\pm$ 0.43 & 2.23 $\pm$ 0.72 & \textbf{1.09 $\pm$ 0.06} \\ 
DDSP-SFX & \textbf{1.55 $\pm$ 0.44}& \textbf{2.03 $\pm$ 0.81} & 1.10 $\pm$ 0.07 \\ 
\hline
\end{tabular}

\caption{Objective audio synthesis performance results.}
\label{tab: similarity}
\end{table}
\normalsize
Referring to Table~\ref{tab: similarity}, our objective measures correlate well, suggesting a significant improvement in impulsive sounds (footsteps and gunshots) after we integrated our transient modelling method. For motor sounds that are steadily pitched across the entire signal, our method performs slightly worse than DDSP. This is expected because introducing the regularization term to the loss poses more challenges to the reconstruction. We also show that our transient modelling method can synthesize sharper attacks for impulsive signals in our accompanying website~\footnote{\url{https://reinliu.github.io/DDSP-SFX/}}.

\subsection{Audio quality}

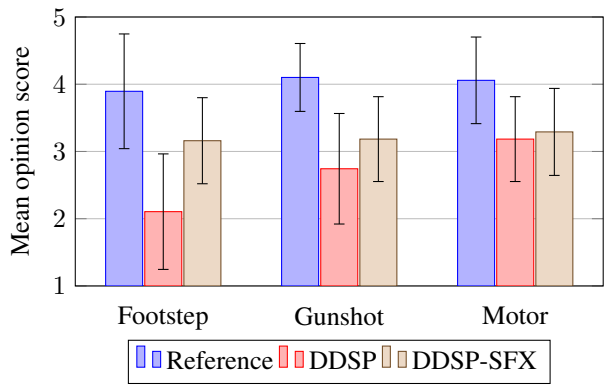
\begin{figure}[htb]
\centering
\begin{tikzpicture}
\label{tab: quality}
    \begin{axis}[
        width  = \linewidth,
        height = 0.6*\linewidth,
        major x tick style = transparent,
        ybar=2*\pgflinewidth,
        bar width=14pt,
        ymin=1, 
        ymax=5,
        ymajorgrids = true,
        ylabel = {Mean opinion score},
        symbolic x coords={Footstep,Gunshot,Motor},
        xtick = data,
        scaled y ticks = false,
        enlarge x limits=0.25,
        legend style={at={(0.5,-0.2)},
        anchor=north,legend columns=-1}
    ]
        \addplot+[error bars/.cd, y dir=both, y explicit, error bar style={black}]
            coordinates {
            (Footstep, 3.895) +- (0.853, 0.853)
            (Gunshot, 4.101) +- (0.505, 0.505)
            (Motor, 4.058) +- (0.644, 0.644)
            };

        \addplot+[error bars/.cd, y dir=both, y explicit, error bar style={black}]
             coordinates {
            (Footstep, 2.105) +- (0.859, 0.859)
            (Gunshot, 2.743) +- (0.822, 0.822)
            (Motor, 3.184) +- (0.631, 0.631)
             };

        \addplot+[error bars/.cd, y dir=both, y explicit, error bar style={black}]
             coordinates {
            (Footstep, 3.160) +- (0.640, 0.640)
            (Gunshot, 3.184) +- (0.631, 0.631)
            (Motor, 3.291) +- (0.647, 0.647)
             };

        \legend{Reference, DDSP, DDSP-SFX}
    \end{axis}
\end{tikzpicture}
\caption{User-rated audio quality of sound effects.}
\label{tab: quality}
\end{figure}

There were 26 participants (M/F: 19/7; ages: 23-53; audio experts/non-experts: 14/12) in our subjective listening test. Each participant was requested to use a pair of headphones for the listening tests. The reference tracks are expected to receive the highest scores(4-5), therefore, we removed two outliers that rated the reference tracks below 2 for over $50\%$ of the questions, resulting in 24 effective participants. Fig.~\ref{tab: quality} shows a bar plot of the collected mean opinion scores for thesound quality with variance as error bars. Our subjective test results are similar to our objective measures, where we see significant improvements for the impulsive sounds. The footsteps generated by DDSP were rated lower than our method as many participants noticed the harmonic artifacts. Further, the motor sounds were rated similarly among the two methods.

\subsection{Timbre encoding}
\label{sec: Timbre encoding}

\begin{table}[ht]
\centering

\begin{tabular}{|c|c|c|c|}
\hline
\multirow{7}{*$z$}{} & Motors & Gunshots & Footsteps \\
\hline
$0.1$ & 0.255 & 0.235 & 0.216\\
$0.5$ & 0.451 & 0.431 & 0.510\\
$1$ & 0.549 & 0.608 & 0.745\\
$2$ & 0.686 & 0.804 & 0.843\\
$3$ & 0.941 & 0.826 & /\\
\hline
\end{tabular}
\caption{Percentage of sounds rated as different from reference.}
\label{tab: encoding}
\end{table}





There were 19 participants (M/F: 11/8; Age: 23-53; Audio expert: 9/10) in our second listening subjective test. Table~\ref{tab: encoding} shows the percentage of the participants who rated the synthesized sounds as different from the reference sounds when we vary the value of $z$. Our subjective test results indicate that most participants recorded a timbre differences for $z > 1$. This follows reasonably because the encoder learns to encode the "most typical" timbres across the data set within one standard deviation. Larger values of $z$ indicate a rarer timbre across the whole data set. Depending on the variations available within the dataset itself, the value of $z$ for which users start to tell the timbre differences will likely change accordingly. Lastly, to demonstrate the effectiveness of our timbre encoding, we show how varying $z$ temporally could contribute to timbre variations in our supplementary website.


\section{Discussion}
\label{sec:page}

In this paper, we integrated DDSP with a transient model and show that it improves the synthesis result for impulsive sound effects. We propose a simple method for controlling the timbre variation of the generated sound effect while also enabling deterministic attribute transfer given a limited dataset. We further demonstrate the out-of-domain timbre transfer capability by using human vocalization as guiding sounds. We hope our method will contribute to creative sound design by allowing users to create realistic sound effects using their own voices as a guiding sound. Future work may include training our model on a larger audio dataset with more variations to enable more expressive timbre control.



\vfill\pagebreak

\bibliographystyle{IEEEbib}
\bibliography{strings,refs,references}

\end{document}